\documentclass[modern]{aastex61}
\usepackage[utf8]{inputenc}
\usepackage{hyperref}
\usepackage{natbib}
\usepackage{graphicx}
\usepackage{xspace}
\usepackage{amsmath}
\usepackage{amssymb}
\usepackage{mathrsfs}

\newcommand{\numax}{\mbox{$\nu_{\rm max}$}\xspace}
\newcommand{\Dnu}{\mbox{$\Delta \nu$}\xspace}

\newcommand{\muHz}{\mbox{$\mu$Hz}\xspace}
\newcommand{\teff}{\mbox{$T_{\rm eff}$}\xspace}

\newcommand{\msun}{\mbox{$\mathrm{M}_{\odot}$}\xspace}
\newcommand{\lsun}{\mbox{$\mathrm{L}_{\odot}$}\xspace}

\newcommand{\kepler}{\emph{Kepler}\xspace}
\newcommand{\hipparcos}{\emph{Hipparcos}\xspace}

\newcommand{\ktwo}{\emph{K2}\xspace}

\newcommand{\twosidedrange}[3]{\ensuremath{#1^{+#2}_{-#3}}}

\newcommand{\HatzesQRange}{\twosidedrange{2.1}{0.7}{0.6}}
\newcommand{\SONGQRange}{\twosidedrange{3.1}{1.2}{0.8}}
\newcommand{\CombinedQRange}{\twosidedrange{2.8}{0.6}{0.5}}

\newcommand{\HatzesNuMaxRange}{\twosidedrange{2.21}{0.13}{0.12}}
\newcommand{\SONGNuMaxRange}{\twosidedrange{2.33}{0.13}{0.14}}

\newcommand{\PeriodRange}{\twosidedrange{629.0}{1.7}{2.1}}

\newcommand{\EccentricityRange}{0.17\pm0.07} 

\newcommand{\PlanetRVAmpRange}{\twosidedrange{127}{15}{14}}

\newcommand{\dd}{\mathrm{d}}
\newcommand{\diff}[2]{\frac{\dd #1}{\dd #2}}

\begin{document}

\title{Aldebaran~b's temperate past uncovered in planet search data}

\author[0000-0003-1540-8562]{Will M. Farr}

\email{w.farr@bham.ac.uk}

\affiliation{Birmingham Institute for Gravitational Wave Astronomy,
  University of Birmingham, Birmingham, B15 2TT, United Kingdom}

\affiliation{School of Physics and Astronomy, University of
  Birmingham, Birmingham, B15 2TT, United Kingdom}

\affiliation{Center for Computational Astrophysics, Flatiron Institute, 162 Fifth Avenue, New York NY 10010, United States}

\author[0000-0003-2595-9114]{Benjamin J. S. Pope}
\affiliation{Sydney Institute for Astronomy (SIfA), School of Physics, University of Sydney, NSW 2006, Australia}
\affiliation{Center for Cosmology and Particle Physics, Department of Physics, New York University, 726 Broadway, New York, NY 10003, USA}
\affiliation{NASA Sagan Fellow}

\author[0000-0002-4290-7351]{Guy R. Davies}

\affiliation{School of Physics and Astronomy, University of
  Birmingham, Birmingham, B15 2TT, United Kingdom}
\affiliation{Stellar Astrophysics Centre, Department of Physics
and Astronomy, Aarhus University, Ny Munkegade 120, DK-8000 Aarhus C, Denmark}

\author{Thomas S. H. North}
\affiliation{School of Physics and Astronomy, University of Birmingham, Birmingham, B15 2TT, United Kingdom}
\affiliation{Stellar Astrophysics Centre, Department of Physics
and Astronomy, Aarhus University, Ny Munkegade 120, DK-8000 Aarhus C, Denmark}

\author[0000-0002-6980-3392]{Timothy R. White}
\affiliation{Stellar Astrophysics Centre, Department of Physics
and Astronomy, Aarhus University, Ny Munkegade 120, DK-8000 Aarhus C, Denmark}

\author{Jim W. Barrett}
\affiliation{Birmingham Institute for Gravitational Wave Astronomy,
  University of Birmingham, Birmingham, B15 2TT, United Kingdom}

\affiliation{School of Physics and Astronomy, University of
  Birmingham, Birmingham, B15 2TT, United Kingdom}

\author{Andrea Miglio}

\affiliation{School of Physics and Astronomy, University of
  Birmingham, Birmingham, B15 2TT, United Kingdom}
\affiliation{Stellar Astrophysics Centre, Department of Physics
and Astronomy, Aarhus University, Ny Munkegade 120, DK-8000 Aarhus C, Denmark}

\author{Mikkel N. Lund}

\affiliation{Stellar Astrophysics Centre, Department of Physics
and Astronomy, Aarhus University, Ny Munkegade 120, DK-8000 Aarhus C, Denmark}
\affiliation{School of Physics and Astronomy, University of
  Birmingham, Birmingham, B15 2TT, United Kingdom}

\author{Victoria Antoci}
\affiliation{Stellar Astrophysics Centre, Department of Physics and Astronomy, Aarhus University, Ny Munkegade 120, DK-8000 Aarhus C, Denmark}

\author{Mads Fredslund Andersen}
\affiliation{Stellar Astrophysics Centre, Department of Physics and Astronomy, Aarhus University, Ny Munkegade 120, DK-8000 Aarhus C, Denmark}

\author{Frank Grundahl}
\affiliation{Stellar Astrophysics Centre, Department of Physics and Astronomy, Aarhus University, Ny Munkegade 120, DK-8000 Aarhus C, Denmark}

\author[0000-0001-8832-4488]{Daniel Huber}
\affiliation{Institute for Astronomy, University of Hawai`i, 2680 Woodlawn Drive, Honolulu, HI 96822, USA}
\affiliation{Sydney Institute for Astronomy (SIfA), School of Physics, University of Sydney, NSW 2006, Australia}
\affiliation{SETI Institute, 189 Bernardo Avenue, Mountain View, CA 94043, USA}
\affiliation{Stellar Astrophysics Centre, Department of Physics and Astronomy, Aarhus University, Ny Munkegade 120, DK-8000 Aarhus C, Denmark}

\begin{abstract}
The nearby red giant Aldebaran is known to host a gas giant planetary companion from decades of ground-based spectroscopic radial velocity measurements. Using Gaussian Process-based Continuous Auto-Regressive Moving Average (CARMA) models, we show that these historic data also contain evidence of acoustic oscillations in the star itself, and verify this result with further dedicated ground-based spectroscopy and space-based photometry with the \kepler Space Telescope. From the frequency of these oscillations we determine the mass of Aldebaran to be $1.16 \pm 0.07$~\msun, and note that this implies its planet will have been subject to insolation comparable to the Earth for some of the star's main sequence lifetime. Our approach to sparse, irregularly sampled time series astronomical observations has the potential to unlock asteroseismic measurements for thousands of stars in archival data, and push to lower-mass planets around red giant stars.
\end{abstract}

\section{Introduction}

Aldebaran ($\alpha$~Tauri) is a well-known first-magnitude red giant star, and has long been the subject of astronomical investigations. It was one of the first stars around which an extrasolar planet candidate was identified, by looking for Doppler shifts from the star's reflex motion around the common centre of mass with its companion \citep[the radial velocity or RV method;][]{struverv}. While the hot Jupiter 51~Peg~b \citep{51peg} was the first exoplanet to be recognized as such, before this, \citet{hatzes1993} had noted RV variations in Pollux \citep[$\beta$~Gem; subsequently confirmed as a planet:][]{betgemconf,betgemconf2}, Arcturus ($\alpha$~Boo, unconfirmed), and Aldebaran. After further investigation by \citet{Hatzes1998}, \citet{Hatzes2015} have now claimed a firm RV detection of a planetary-mass companion Aldebaran~b, with a period of $629.96 \pm 0.90$~d.

In this paper, we present a re-analysis of these original RV data in which we not only confirm the planetary signal, but detect acoustic oscillations in Aldebaran for the first time. \citet{hatzes1993} noted night-to-night RV variability: this is the noise floor limiting the sensitivity to sub-$M_J$ planets around giants in RV surveys \citep{2005PASJ...57...97S,2014A&A...566A.113J}. We pull out the asteroseismic signal in this planet hunting noise. We validate this method and its result with new independent RV observations with the Hertzsprung SONG~Telescope, and photometry from the \ktwo Mission. By measuring the frequency of maximum power of these \emph{p}-mode oscillations, \numax, we asteroseismically determine the mass of Aldebaran to be $1.16 \pm 0.07$~\msun. This precise stellar mass allows us to calculate that Aldebaran~b and any satellites it may have, although they are now likely to be very hot, would have had equilibrium temperatures comparable to that of the Earth when Aldebaran was on the main sequence. It is possible that they may have once been habitable, billions of years ago.

Our new approach to asteroseismic data analysis, based on Continuous Auto-Regressive Moving Average (CARMA) models, can extract exoplanet signals together with measures of \numax from sparse and irregularly-sampled time series. An all-sky survey to find planetary companions and to precisely measure the masses of all nearby red giant stars is feasible with this new approach, and the required data either already exist in large radial velocity exoplanet surveys, or are easy to obtain with ground-based telescopes.

\section{Asteroseismology of Red Giants}
Asteroseismology is a powerful tool for the charaterisation of red giant stars \citep[see][for detailed reviews]{2013ARA&A..51..353C, hekker17}.  Red giants exhibit oscillations that are excited and damped by stellar convection.  In the power spectrum of either radial velocity or photometric observations, these modes of oscillation form a characteristic pattern of peaks which can be used to infer intrinsic stellar properties \citep[e.g.,][]{2016AN....337..774D}.  The easiest properties of the pattern to determine are the frequency of maximum amplitude \numax and the so-called large separation \Dnu \citep{Kjeldsen95}, which are often referred to as global asteroseismic parameters.

These parameters can be used to estimate the radius $R$, mass $M$, and surface gravity $g$ of stars when combined with an estimate of the effective temperature \teff\ through scaling relations:

\begin{eqnarray}
\left( \frac{R}{\mathrm{R_{\odot}}} \right) \simeq & \left( \frac{\nu_{\rm max}}{\nu_{\rm max, \odot}} \right) \,
\left( \frac{\Delta \nu}{\Delta \nu_{\odot}} \right)^{-2} \, \left( \frac{T_{\rm eff}}{T_{\rm eff, \odot}} \right)^{0.5}, \\
\left( \frac{M}{\mathrm{M_{\odot}}} \right) \simeq & \left( \frac{\nu_{\rm max}}{\nu_{\rm max, \odot}} \right)^{3} \,
\left( \frac{\Delta \nu}{\Delta \nu_{\odot}} \right)^{-4} \, \left( \frac{T_{\rm eff}}{T_{\rm eff, \odot}} \right)^{1.5}, \\
\left( \frac{g}{\mathrm{g_{\odot}}} \right) \simeq & \left( \frac{\nu_{\rm max}}{\nu_{\rm max, \odot}} \right) \,
 \left( \frac{T_{\rm eff}}{T_{\rm eff, \odot}} \right)^{0.5}.
\end{eqnarray}

While the accuracy of these scaling relations is still a matter of ongoing work \citep[e.g.,][]{2017ApJ...844..102H}, stellar parameters can be estimated by comparing observables to parameters from models of stellar evolution.

The ability to measure \numax and \Dnu depends on the length and sampling rate of a data set.
In practice, for the higher luminosity red giants it is more straightforward to measure \numax than \Dnu.  Typical values for \numax\ range from $0.1-20 \rm\,\mu Hz$ for luminous giants and $20-50 \rm\,\mu Hz$ for stars near the red clump.  For the evolutionary stages, \Dnu typically ranges from $0.02-3 \rm\,\mu Hz$ and $3-7 \rm\,\mu Hz$ depending on stellar mass and \teff
\citep[e.g.,][]{2011A&A...525L...9M, 2013A&A...559A.137M}.  Hence, \numax requires less frequency resolution than \Dnu to establish a good measurement.


\section{Time-Domain Models}

Radial velocity variations intrinsic to the star, such as those caused by stellar oscillations, have previously limited the precision with which red giant planets have been studied \citep{2005PASJ...57...97S}. We aim to model these noise processes and use them to recover asteroseismic information, as well as to improve our estimates of the orbital parameters of the planet. It is easy to observe Aldebaran and similarly-bright stars with ground-based spectroscopic instruments, typically only requiring short exposures that can be obtained even under adverse observing conditions. There is indeed a considerable archive of such observations already, as a legacy of radial velocity (RV) surveys conducted to find exoplanets. In most cases, however, these have not so far been useful for asteroseismology because these RV data are sparsely and irregularly sampled. Because we have to pause observations during the day, during poor weather conditions, or simply when targets of higher priority are being observed, we get time series which may have significant and uneven gaps. This introduces a window-function effect: the power spectrum as constructed for example by a Fourier transform, or a Lomb-Scargle Periodogram \citep{lomb,scargle}, is convolved with the Fourier transform of the window function, introducing strong sidelobes adjacent to real frequency peaks and causing crosstalk between adjacent frequency channels. This imposes significant limitations both on the signal-to-noise and frequency resolution of power spectra derived from linear methods such as the Lomb-Scargle periodogram, and in practice makes asteroseismology with these conventional approaches difficult or impossible from the ground for stars with oscillation periods ranging from $\sim 12$~h to $\sim$~a few days.

If we apply nonlinear statistical inference methods this situation can be
improved. \citet{brewer2009} show that a system of driven, damped harmonic
oscillators such as we encounter in asteroseismology can be statistically
modelled as a Gaussian Process (GP), with a covariance kernel consisting of a sum of
damped sinusoids. The hyperparameters of this process encode features of the
power spectral density distributions and are insensitive to the window function. GPs with quasi-periodic kernels have previously been used to re-analyse RV data for main sequence stars \citep{2014MNRAS.443.2517H,2015MNRAS.452.2269R}, where the noise is from stellar activity, but have not previously been applied to red giant asteroseismology.
Unfortunately, these methods are impractical for long time series, as the computational
cost of evaluating the standard GP likelihood function scales as
$\mathcal{O}(n^3)$.

Fortunately, the damped and driven harmonic oscillator GP can be written as the
solution of a class of stochastic ordinary differential equations, the
Continuous Auto-Regressive Moving Average (CARMA) models, whose likelihoods can
be evaluated in linear time. For problems such as this, we therefore have access
to these powerful computational tools for inferring their power spectral
densities.  Our treatment of CARMA models is described more fully in Appendix
\ref{carmatheory}.   \citet{Kelly2014} also give more details on CARMA
processes in an astronomical context, and show how a state-space model of the
process $y(t)$ can be tracked through a time-series of uncertain observations
$y_k = y\left( t_k \right) + \epsilon_k$, with Gaussian noise $\epsilon_k$ of
zero mean and known (heteroskedastic) variance, using a Kalman filter to produce
a likelihood function $p\left( y_k \mid \sigma, r_i, b_j \right)$ depending on
the amplitude of the stochastic forcing, $\sigma$, the (possibly complex)
eigenfrequencies of the ODE, $r_i$, and parameters describing the power spectrum
of the stochastic forcing, $b_j$ in $\mathcal{O}\left( n \right)$ computational
time. \citet{Foreman-Mackey2017} has demonstrated that the same model can also be implemented via a novel matrix factorization, and demonstrated its asteroseismic potential modelling the light curve of KIC~11615890.

Here we have adopted the \citet{Kelly2014} approach.  The \texttt{CARMA.jl}
package\footnote{\url{https://github.com/farr/CARMA.jl}} implements Kalman
filters that can compute the likelihood for a set of observations parameterised
by either the $r_i$ and $b_j$ or the RMS amplitudes $A_i$ of each eigenmode of
the ODE and the corresponding roots $r_i$. This likelihood is computed for the
residuals of a deterministic Keplerian RV mean model for the planetary
companion.  Because we have a set of observations taken with different
instruments at different sites (see Section \ref{sec:archival-obs}), we include
RV offset (i.e.\ mean value) parameters that are instrument- and site-dependent
and also an instrument- and site-dependent uncertainty scaling parameter.  Thus,
the $i$th RV measured at site and intstrument combination $j$, $y_{ij}$, is
assumed to be
\begin{equation}
  \label{eq:RV-model}
  y_{ij} = v_r\left(t_i\right) + \mu_j + y\left( t_i \right) + \nu_j
\epsilon_{ij},
\end{equation}
where $t_i$ is the time of the measurement, $v_r$ is the planetary RV signal,
$\mu_j$ is the mean RV offset for site and instrument combination $j$, $y(t)$ is
the CARMA process representing stellar activity, $\nu_j$ is the uncertainty
scaling parameter for site and instrument combination $j$, and $\epsilon_{ij}$
is an independent Gaussian random variable with mean zero and standard deviation
equal to the quoted uncertainty of the observation.  We impose broad priors (see
Table \ref{tab:parameters}) on the parameters of the planetary RV signal, the
CARMA process, the $\mu_j$ offsets, and the $\nu_j$ scale factors.  We treat the
frequency of maximum asteroseismic power, $\nu_\mathrm{max}$, as the frequency
of an appropriate eigenmode in the CARMA process representing the stellar
activity; this is equivalent to fitting a Lorentzian profile to a set of
unresolved asteroseismic modes.  We used the nested sampling algorithm of the
\texttt{Ensemble.jl} package\footnote{\url{https://github.com/farr/Ensemble.jl};
this package implements several stochastic sampling algorithms based on the
``stretch move'' proposal used in \texttt{emcee} \citep{Foreman-Mackey2013} and
introduced in \citet{Goodman2010}} to calculate the marginal likelihood
(evidences) and draw samples from the posterior distribution over the parameters
of our various models.

\section{Aldebaran}
Aldebaran is a red giant star with spectral type K5, one of the nearest such stars at a distance of only $19.96 \pm 0.38$~pc as determined by \hipparcos \citep{hipparcos}. Its position near the Ecliptic permits the determination of its angular diameter by lunar occultations and by interferometry \citep[$20.58 \pm 0.03$ mas;][]{richichi2005,1979ApJ...228L.111B,brown1979,panek1980}. These tight constraints are valuable in breaking degeneracies in stellar modelling and make this an ideal star for asteroseismic characterization (Appendix~\ref{sec:mod}).

\subsection{Archival Observations}
\label{sec:archival-obs}

\begin{figure}
\centering
\plotone{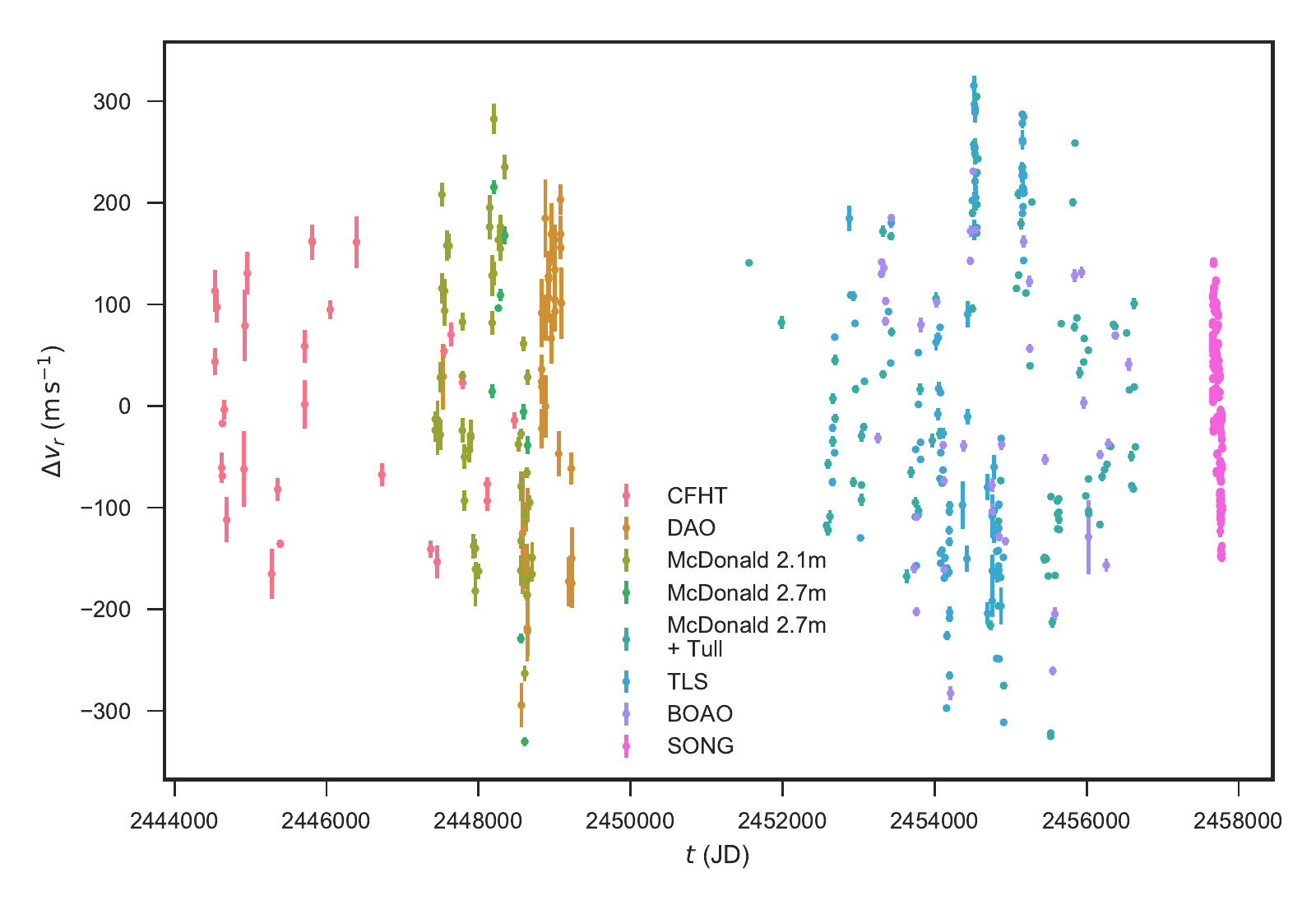}
\caption{All archival and new radial velocities used for inference in this work.  The site and instrument for each data set is indicated in the legend (see text for a full description).}
\label{alldata}
\end{figure}

\begin{figure}
  \includegraphics[width=\columnwidth]{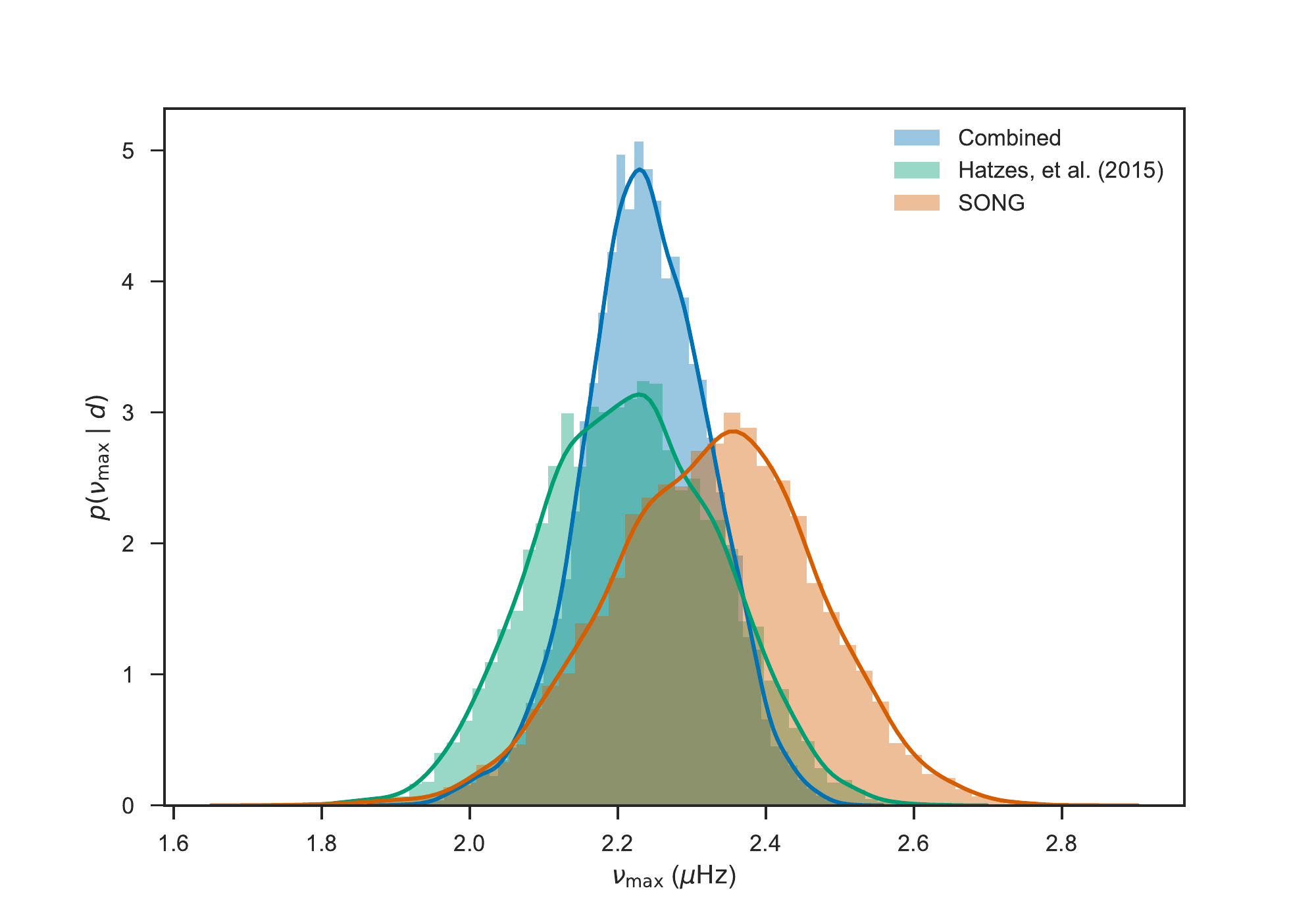}
  \caption{Posterior for \numax{} from the complete, combined RV data set described in this paper (blue), the archival data from \citet{Hatzes2015} (green), and the SONG data described in Section \ref{sec:SONG-observations} (red).  The inference on the archival data is consistent with the independent inference on the SONG data set alone.}
  \label{fig:numax-datasets}
\end{figure}

\citet{Hatzes2015} reported on RV observations of Aldebaran from the coud\'{e}
\'{e}chelle spectrograph of the Th\"{u}ringer Landessternwarte Tautenburg (TLS),
the Tull Spectrograph of the McDonald 2.7 m telescope, and the Bohyunsan
Observatory \'{E}chelle Spectrograph (BOES) spectrograph of the Bohyunsan
Optical Astronomy Observatory (BOAO) spanning the period from 2000.01 to
2013.92.  Combined with earlier observations \citep{hatzes1993}, we construct a
full time series spanning more than two decades from 1980.80 to 2013.71
(Figure~\ref{alldata}) and fit this with a Keplerian model and a CARMA process.
We find evidence for a low-quality ($Q = \HatzesQRange{}$) oscillatory mode with
$\nu_\mathrm{max} = \HatzesNuMaxRange{} \, \mu\mathrm{Hz}$ (here and throughout
we quote the posterior median and the range from the 0.16 to the 0.84 posterior
quantile).  This is consistent with the presence of a number of un-resolved
asteroseismic modes in the RV data.  The low Q factor in this case refers to the
inverse fractional width of the band over which the modes are excited, not the
quality of any individual mode.  Our posterior for \numax{} from this data set
is shown in Figure \ref{fig:numax-datasets}.

\subsection{SONG Observations}    
\label{sec:SONG-observations} 
\begin{figure}
\centering
\includegraphics[width=\textwidth]{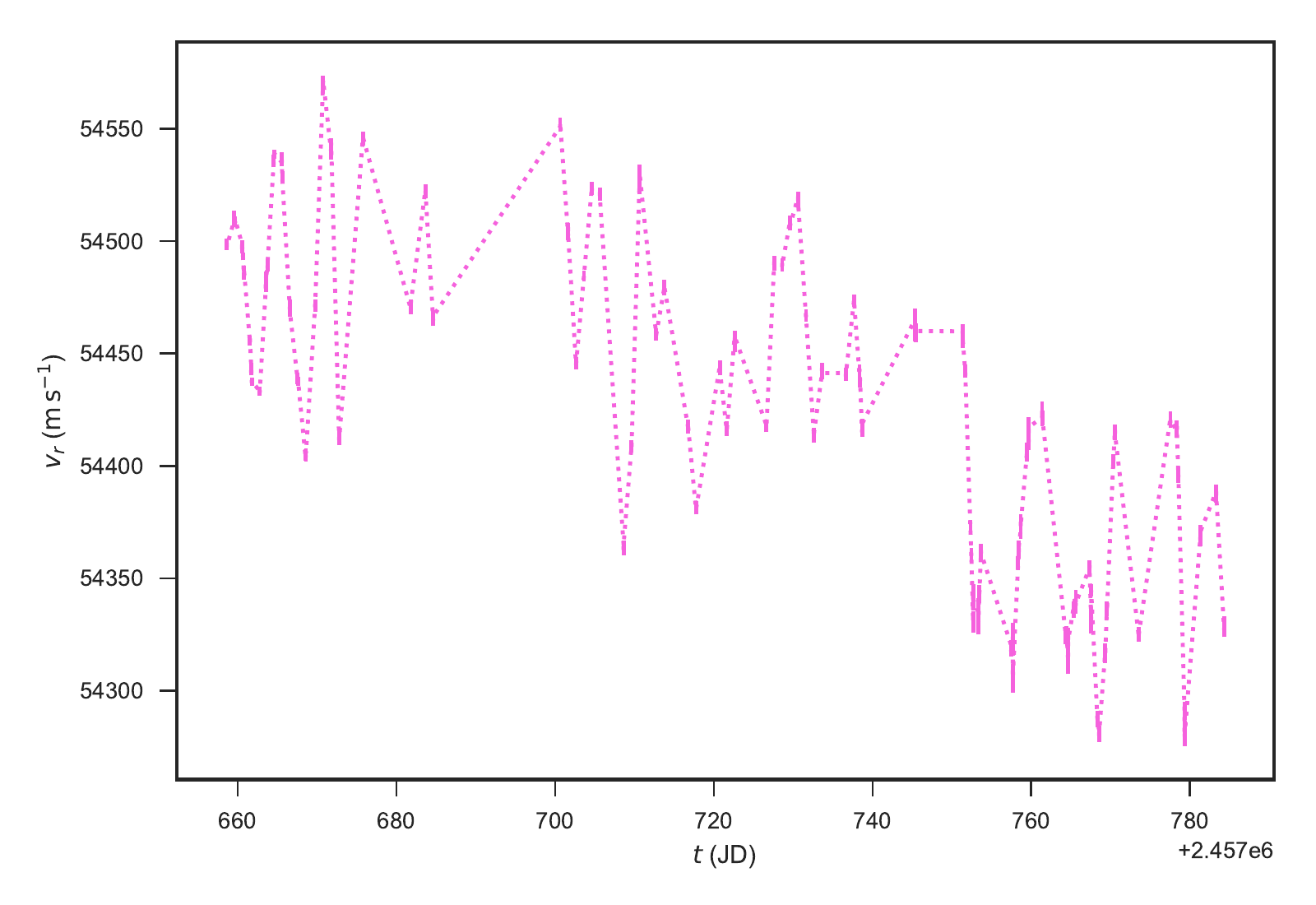}
\caption{New radial velocities of Aldebaran obtained by the Hertzsprung~SONG Telescope. Clear oscillations can be seen, plus a trend representing part of the planetary orbit.}
\label{songdata}
\end{figure}

In order to confirm the oscillations detected in
the archival data, we used the Hertzsprung SONG telescope
\citep{2017ApJ...836..142G} to conduct high time cadence follow-up observations.
These were carried out in the highest resolution mode ($R\approx110000$) with an
integration time of 30\,s, between 2016 September 27 and 2017 January 12. During
this time we attempted to obtain at least one visit per available night, for a
total of 254 epochs over the campaign. The spectra were extracted using the SONG
pipeline \citep[see][]{2017ApJ...836..142G}. SONG employs an iodine cell for
precise wavelength calibration and the velocities were determined using the
\textit{iSONG} software following the same procedures as described in
\cite{2017ApJ...836..142G}. The typical velocity precision per visit is
2.5$\,\rm m\,s^{-1}$ allowing us to easily detect the oscillations which display
a peak-to-peak amplitude of $\approx \, 170 \, \rm m \, s^{-1}$.
Figure~\ref{songdata} displays the radial velocities obtained---oscillations
are easily visible as well as a long-term trend, due to the planetary companion.
From this data set alone we also find evidence for a low-quality ($Q =
\SONGQRange{}$) oscillatory mode with $\nu_\mathrm{max} = \SONGNuMaxRange{} \,
\mu\mathrm{Hz}$ which is consistent with the inference from the archival data.
This posterior is shown in Figure \ref{fig:numax-datasets}.

\begin{figure}
\centering
\includegraphics[width=\textwidth]{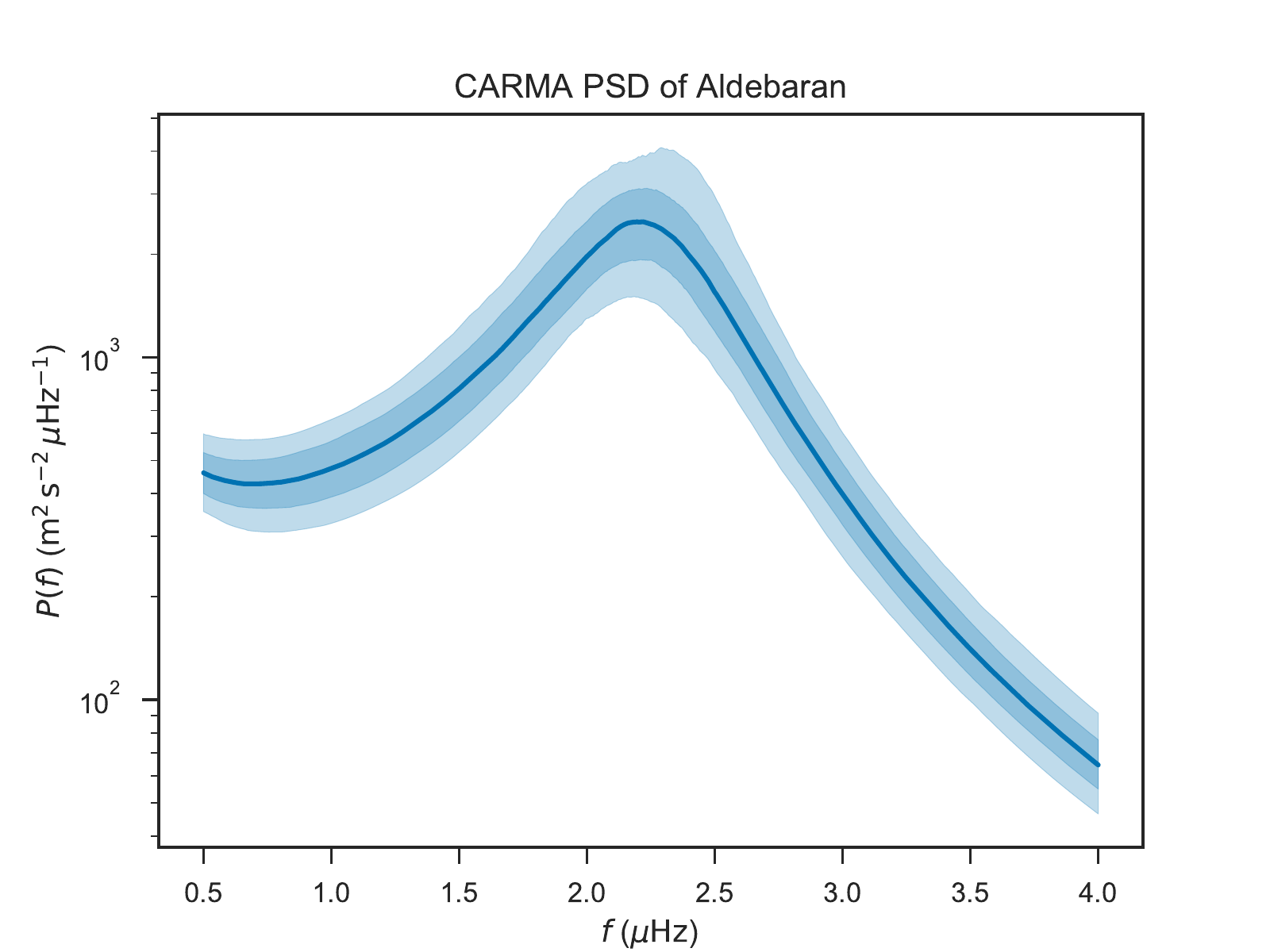}
\caption{CARMA-derived power spectral density from archival and SONG RV observations of Aldebaran, showing the clear peak at around 2\,\muHz. The dark and light shaded regions represent $1\,\sigma$~and $2\,\sigma$ posterior probability contours. }
\label{psd}
\end{figure}

\subsection{K2 Observations}

In order to verify the results of the novel analysis presented above, we sought to obtain an independent detection of the oscillations of Aldebaran and compare the frequencies determined with the two methods.
Aldebaran was observed with \ktwo \citep{howell14}, a two-wheeled revival of the \kepler Mission \citep{2010sci...327..977b}, under Guest Observer Program 130471 in Campaign~13, from 2017 March 8 to 2017 May 27. Aldebaran and similar high-luminosity K giants are well known to show photometric variability due to solar-like oscillations \citep{bedding2000}, and hence the continuous, high-precision K2 light curves allow an independent confirmation of our CARMA results.

As Aldebaran is extremely bright, it saturates the \kepler detector and it is therefore not possible to use standard photometry pipelines to extract a K2 lightcurve. We instead use halo photometry \citep{White2017}, whereby unsaturated pixels from the outer part of the large and complicated halo of scattered light around bright stars are used to reconstruct a light curve as a weighted linear combination of individual pixel-level time series, as described in Appendix~\ref{halo}.

Figure \ref{k2_lightcurve} shows the light curve and power spectrum of the K2 observations of Aldebaran. We detect clear variability on an average timescale of $\sim$\,5\,days, consistent with the expected timescale for solar-like oscillations for a high-luminosity red giant. To measure global asteroseismic parameters we model the background variability in the Fourier domain using the methodology described in \citet{huber09}, yielding a frequency of maximum power of $\numax = 2.2 \pm 0.25\,\muHz$ and amplitude per radial mode of $1850 \pm 500$\,ppm. Due to the limited frequency resolution of the 70-day K2 time series we were unable to measure the large frequency separation \Dnu, which is expected to be $\approx$\,0.5\muHz.

Using the \textsc{k2ps} planet-search code \citep{k2ps,Pope2016} to examine this light curve, we search for transits across a wide range of periods, and find no evidence either of short-period planetary transits or an eclipsing stellar companion.

\begin{figure}
\centering
\includegraphics[width=\textwidth]{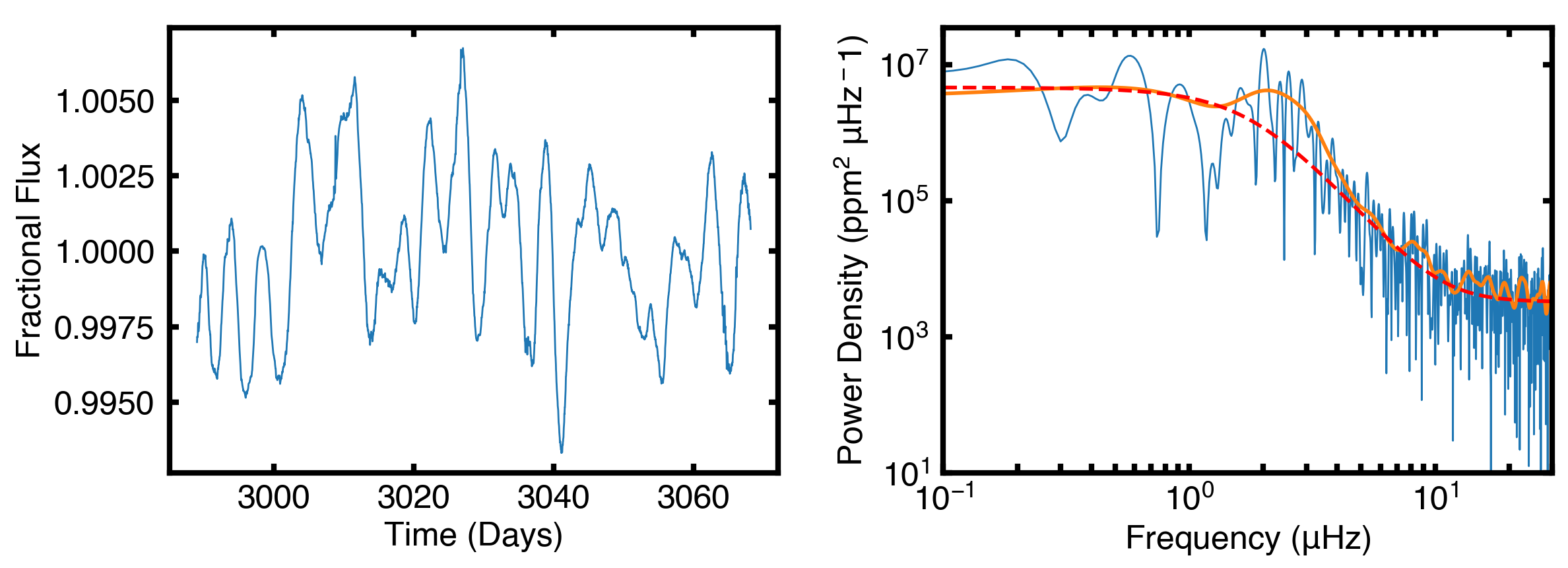}
\caption{\ktwo lightcurve (left) and power spectrum (right) of Aldebaran. The red dashed line shows the background model, and the orange line is a heavily smoothed version of the power spectrum used to measure the frequency of maximum power. }
\label{k2_lightcurve}
\end{figure}

\subsection{Planetary and Stellar Parameters}
\label{sec:planet-star-parameters}

By jointly fitting a Keplerian for the planet with a CARMA model for the stellar
oscillations to the combined archival and SONG data sets (Figure \ref{alldata}),
we obtain a precise estimate of both \numax and the planetary orbital
parameters. The planetary orbital parameters we obtain are similar to
\citet{Hatzes2015}, but with larger uncertainty that is likely due to our
more-flexible and correlated model for the stellar component of the RV signal:
period $P = \PeriodRange{} \, \mathrm{d}$,  eccentricity $e =
\EccentricityRange{}$; and radial velocity semi-amplitude $K =
\PlanetRVAmpRange{} \, \mathrm{m}\, \mathrm{s}^{-1}$.

The stellar oscillations are best fit with a single, low-quality ($Q =
\CombinedQRange{}$) oscillatory mode with $\numax = 2.24^{+ 0.09}_{- 0.08}$
\muHz.  We find no improvement in the marginal likelihood (evidence) or other
model-selection information criteria \citep{Gelman2013} from models with
additional oscillatory modes, so we conclude that the large spacing $\Delta\nu$
is not observable in this data set.

We have used the additional constraint of \numax determined from this analysis
to update the stellar properties of Aldebaran.  The details of the stellar
modelling are presented in Appendix \ref{sec:mod}.  We have considered the
impact of our additional \numax constraint and run our analysis for different
sets of spectroscopic estimates.  For final values we adopt the
\citet{2012Sheffield} spectroscopic solution as being `middle of the road'
estimates.  For the analysis without constraint on \numax we find estimates of
the stellar properties as $M = 1.27^{+0.24}_{-0.20} \, \mathrm{M_{\odot}}$ and
age $4.9^{+3.6}_{-2.0} \, \rm Gyr$.  With the inclusion of \numax we find $M =
1.16\, \pm \, {0.07} \, \mathrm{M_{\odot}}$ and age $6.4^{+1.4}_{-1.1} \, \rm
Gyr$.  It is clear that the addition of \numax provides substantially more
precise estimates of mass and age. With this stellar mass, the radial velocity
signal translates to an inclination-dependent planetary mass of $M\,\sin\,{i} =
5.8 \pm 0.7~M_J$ \citep[Equation~1]{Torres2008}, a massive giant planet; for
sufficiently high inclinations ($\gtrsim 45^\circ$), the mass of Aldebaran~b may
exceed $13~M_J$, making it a possible low-mass brown dwarf candidate. A recent
study by \citet{2018arXiv180105239H} has cast doubt on the validity of the
detection of the massive planet $\gamma$~Draconis~b, and by extension similar
planets around similar K~giants such as Aldebaran. We remain convinced that the
$\gamma$~Draconis problem is not an issue in this case, and Aldebaran~b is a
bona fide planet, for reasons outlined in Appendix~\ref{gamma_dra}.

\begin{figure}
\centering
\includegraphics[width=\textwidth]{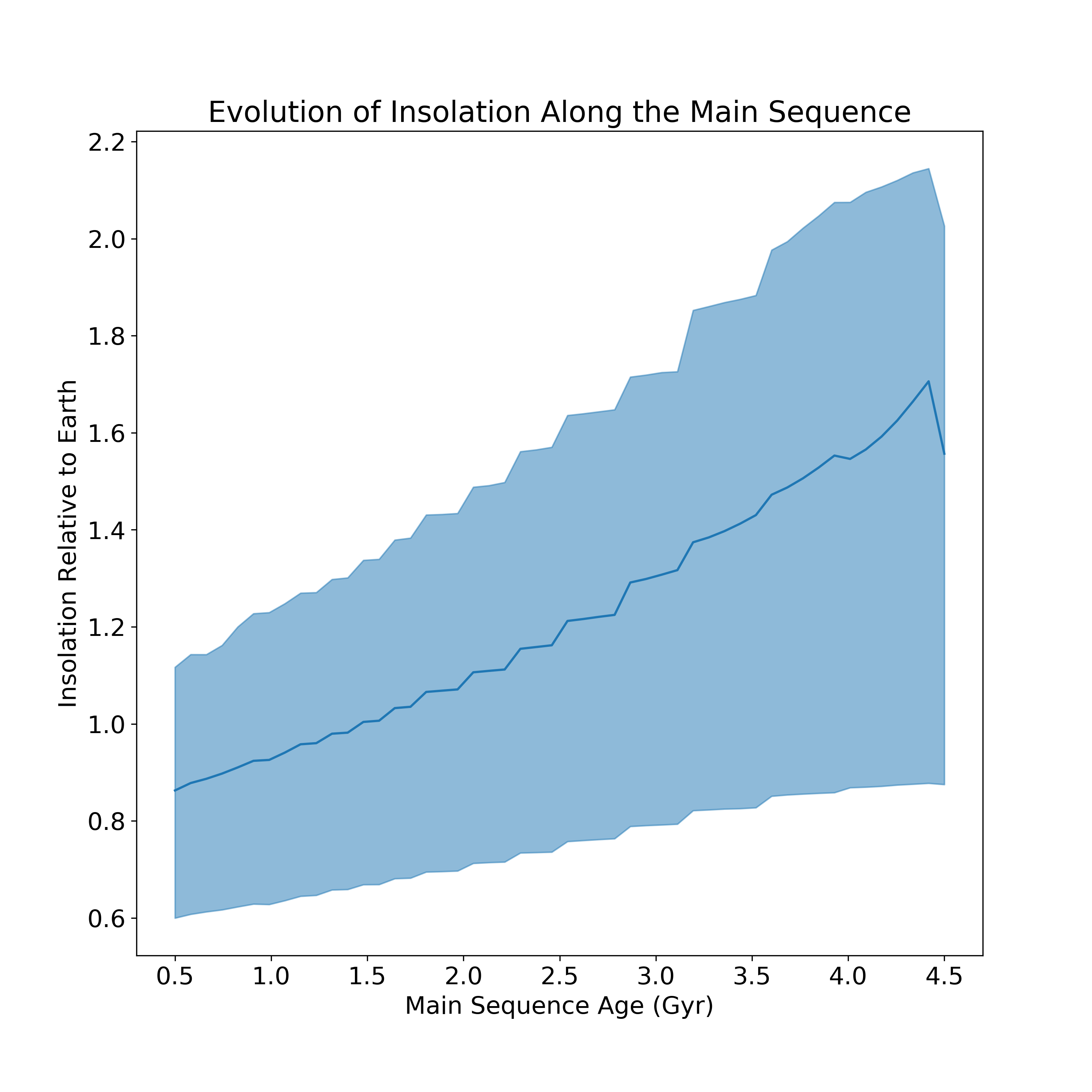}
\caption{Evolution of the insolation of Aldebaran~b relative to the incident sunlight on Earth, as a function of the age of Aldebaran on the main sequence. The solid line is the mean of simulations, and the shaded region represents $1\,\sigma$ equivalent deviations. It is likely that Aldebaran~b spent some time on the main sequence with an equilibrium temperature similar to Earth.}
\label{insolation}
\end{figure}

With the determination of the stellar mass it is possible to infer the parameters Aldebaran had while it was on the main sequence. We conduct a Monte Carlo simulation, drawing masses randomly from a Gaussian distribution $M \sim \mathcal{N}(1.16,0.07)$ and metallicities $[\text{Fe}/\text{H}] \sim \mathcal{N}(-0.15,0.2)$ \citep{decin2003} to predict the luminosity of the main-sequence progenitor, and the semi-major axis of the planet's orbit.
Using the Mesa Isochrones and Stellar Tracks \citep[MIST:][]{mist0,mist1} models, we compute the evolution of the luminosity $L$ of the star along the main sequence; $L$ evolves from $2.0 \pm 0.7$~\lsun at 0.5~Gyr to $3.5 \pm 2.3$~\lsun at 4.5~Gyr.
We note that this implies that the planet at a semi-major axis of $1.50\, \pm\, 0.03$~AU would have been subject to an insolation comparable to that of the Earth, evolving from 0.5~Gyr to 4.5~Gyr from $0.86\, \pm\, 0.27$ to $1.55 \,\pm\,1.0$ times that of the Earth today (Figure~\ref{insolation}). Subject as well as this to the great uncertainties of the planet's orbital evolution and albedo, Aldebaran~b and any of its moons (or its S-type planets if it is a brown dwarf) may well have hosted temperate environments for some of their history, now long-since destroyed by their star's evolution away from the main sequence.

\section{Conclusions}

Using sophisticated time-domain CARMA models for the stellar activity in
Aldebaran, we have confirmed the previously-suspected planet and additionally
detected acoustic oscillations that permit a mass determination with a precision
of $\sim 6\%$. We have confirmed both of these results with new ground-based
observations, and space-based photometry with \ktwo.

From this pilot study with Aldebaran, we have shown that with limited quantities of data of limited quality, we can nevertheless do asteroseismology from the ground and obtain very precise estimates of stellar parameters. Furthermore, improvements in the algorithm may also permit the detection of not only~\numax but also~\Dnu from similar measurements, permitting more sophisticated asteroseismic analysis. Sufficient radial velocity data either already exist, or are trivial to obtain, in order to do this for essentially all bright giants; while for asteroseismology of solar-like stars, these new methods will allow significantly relaxed observing requirements. One rich archive is the series of observations with the Hamilton \'{E}chelle Spectrograph at Lick Observatory, studying 373 G~and K~giants \citep[e.g.][]{frink2001,hekker2006,ortiz2016}. The method may also be useful in detecting rotational or other periodic variability in the LCES HIRES/Keck Precision Radial Velocity Exoplanet Survey \citep{butler17} of 1,624 FGKM dwarfs, although the time sampling is by design too coarse to detect $p$-mode oscillations in these stars. Furthermore, CARMA models and related methods will enhance deep, all-sky, sparse photometric surveys: an immediate future test of this will be from \hipparcos; while only~58 epochs of photometry are available for Aldebaran, at a sampling we find to be insufficient for our purposes, stars at higher latitudes may often have 150--200 epochs and more even sampling \citep{hipparcos_phot}, and the~14 bright~K giants in \hipparcos noted by \citet{bedding2000} are an ideal first test case. This can naturally be extended to \emph{Gaia} in space \citep{gaia}, or \emph{LSST} from the ground \citep{dmt,lsst,lsstbook}, from which many thousands of new asteroseismic determinations will be possible. In future RV surveys, it may be possible to beat the \citet{2005PASJ...57...97S} $\sim 30$~m/s RV precision limit for red giants by taking many closely spaced observations and modelling-out the effects of stellar oscillations: this will allow us to dig deeper into this intrinsic stellar noise to detect less-massive planets around these stars.

\acknowledgments

We would like to thank Tim Bedding, David Hogg and Dan Foreman-Mackey for their helpful discussions.

We would like to commend the authors of \citet{Hatzes2015} for making their data
publicly available on SIMBAD.

The authors would like to acknowledge the Gadigal people of the Eora Nation on whose ancestral lands the University of Sydney is built.

BP is grateful for funding from the Breakthrough Prize Foundation. This work was performed in part under contract with the Jet Propulsion Laboratory (JPL) funded by NASA through the Sagan Fellowship Program executed by the NASA Exoplanet Science Institute.
TRW acknowledges the support of the Villum Foundation (research grant 10118).
D.H. acknowledges support by the Australian Research Council's Discovery Projects funding scheme (project number DE140101364) and support by the National Aeronautics and Space Administration under Grants NNX14AB92G and NNX16AH45G issued through the Kepler Participating Scientist Program and the K2 Guest Observer Program.

Funding for the Stellar Astrophysics Centre is provided by the Danish National Research Foundation.
This research has made use of the SIMBAD database, operated at CDS, Strasbourg, France, and NASA's Astrophysics Data System. The \kepler data presented in this paper were obtained from the Mikulski Archive for Space Telescopes (MAST). STScI is operated by the Association of Universities for Research in Astronomy, Inc., under NASA contract NAS5-26555. Support for MAST for non-HST data is provided by the NASA Office of Space Science via grant NNX13AC07G and by other grants and contracts.

This work has made use of observations made with the Hertzsprung SONG telescope operated on the Spanish Observatorio
  del Teide on the island of Tenerife by Aarhus and Copenhagen Universities and by the Instituto de
  Astrof\`{i}sica de Canarias.

The Villum Foundation, the Danish Council for Independent Research~|~Natural Sciences and
 the Carlsberg Foundation are thanked for their support on building the SONG prototype on
 Tenerife.
 We thank Jens Jessen-Hansen, Rasmus Handberg and René Tronsgaard Rasmussen for their contribution
 to the development of the SONG data reduction pipeline and archive system.

 WMF, GRD, TN, JWB, AM, and MNL are supported in part by the STFC.

\software{\textsc{IPython} \citep{PER-GRA:2007}, SciPy \citep{scipy}, \textsc{matplotlib} \citep{Hunter:2007}, \textsc{Astropy} \citep{2013A&A...558A..33A}, \textsc{NumPy} \citep{van2011numpy}, \textsc{isochrones} \citep{isochrones}, \textsc{Julia} \citep{Bezanson2017}.}

The data, code, and \LaTeX{} source used to produce this document can be found
at \url{https://github.com/farr/Aldebaran} under an open-source ``MIT'' license.


\appendix

\section{CARMA Models}
\label{carmatheory}

A CARMA process corresponds to the solution, $y(t)$, of the stochastic ODE
\begin{equation}
  \label{eq:GP-definition}
  \prod_{i=1}^{p} \left[ \diff{}{t} - r_i \right] y(t) =
  \prod_{j = 1}^{q} \left[ \diff{}{t} - b_i \right] \eta(t),
\end{equation}
where $p, q \in \mathbb{Z}$, $p > q \geq 0$ define the order of the process;
$r_i \in \mathbb{C}$, $\Re r_i < 0$ are the roots of the characteristic equation
of the ODE; $b_j \in \mathbb{C}$, $\Re b_j < 0$ are corresponding roots of the
inverse process; and $\eta(t)$ is a white-noise GP with
$\left\langle \eta(t) \right\rangle = 0$ and $\left\langle \eta(t) \eta(t')
\right\rangle = \sigma^2 \delta\left(t - t' \right)$.  The constraint that $\Re
r_i < 0$ and $\Re b_j < 0$ ensures that the linear operator defining the process
and its inverse are invertable.  Under the assumption that the solution, $y(t)$,
is real, roots $r_i$ and $b_j$ must either be real or occur in complex conjugate
pairs.  The power spectrum of the process is
\begin{equation}
  \label{eq:CARMA-PSD}
  P_y(f) = \sigma^2 \left| \frac{\prod_{j=1}^q \left[ 2 \pi i f - b_j \right]}{\prod_{i=1}^p \left[ 2 \pi i f - r_i \right]} \right|^2.
\end{equation}
The power spectrum is a rational function of frequency with poles at $2\pi i f =
r_i$ and zeros at $2\pi i f = b_i$; because of the invertable constraints, these
poles and zeros all occur when $\Im f \neq 0$.  The autocorrelation function
corresponding to the power spectrum in Eq.\ \eqref{eq:CARMA-PSD} is
\begin{equation}
  \label{eq:CARMA-ACF}
  \rho_y (\tau) = \left\langle y\left( t \right) y \left( t + \tau \right) \right\rangle = \sum_{i=1}^p A_i^2 e^{r_i \tau},
\end{equation}
where the RMS amplitudes of the $p$ different modes, $A_i$, are functions of
$\sigma$ and the values of the roots $r_i$ and $b_j$.  If $q = p-1$, then the
$A_i$ are independent, while if $q < p-1$ then there are interdependencies among
the $A_i$.  From either the power spectrum in Eq.\ \eqref{eq:CARMA-PSD} or the
autocorrelation function in Eq.\ \eqref{eq:CARMA-ACF}, it is apparent that a
real root $r_i$ represents an exponentially-decaying mode with $r_i$ the rate
constant, while a complex conjugate pair of $r_i$ represent an oscillatory mode
with decay rate $\Re r_i$ and angular frequency $\left|\Im r_i\right|$.
Oscillatory modes can also be described by their frequency $f$ and quality
factor $Q$ via $r = f/(2Q) \pm 2\pi i f$.  Decaying modes generate a damped
random walk, also known as a Ornstein-Uhlenbeck process.  More discussion of CARMA processes, including formulae for $A_i$ in terms of $\sigma$, $r_i$, and $b_j$, can be found in \citet{Kelly2014} and references therein.

\begin{figure}
\plotone{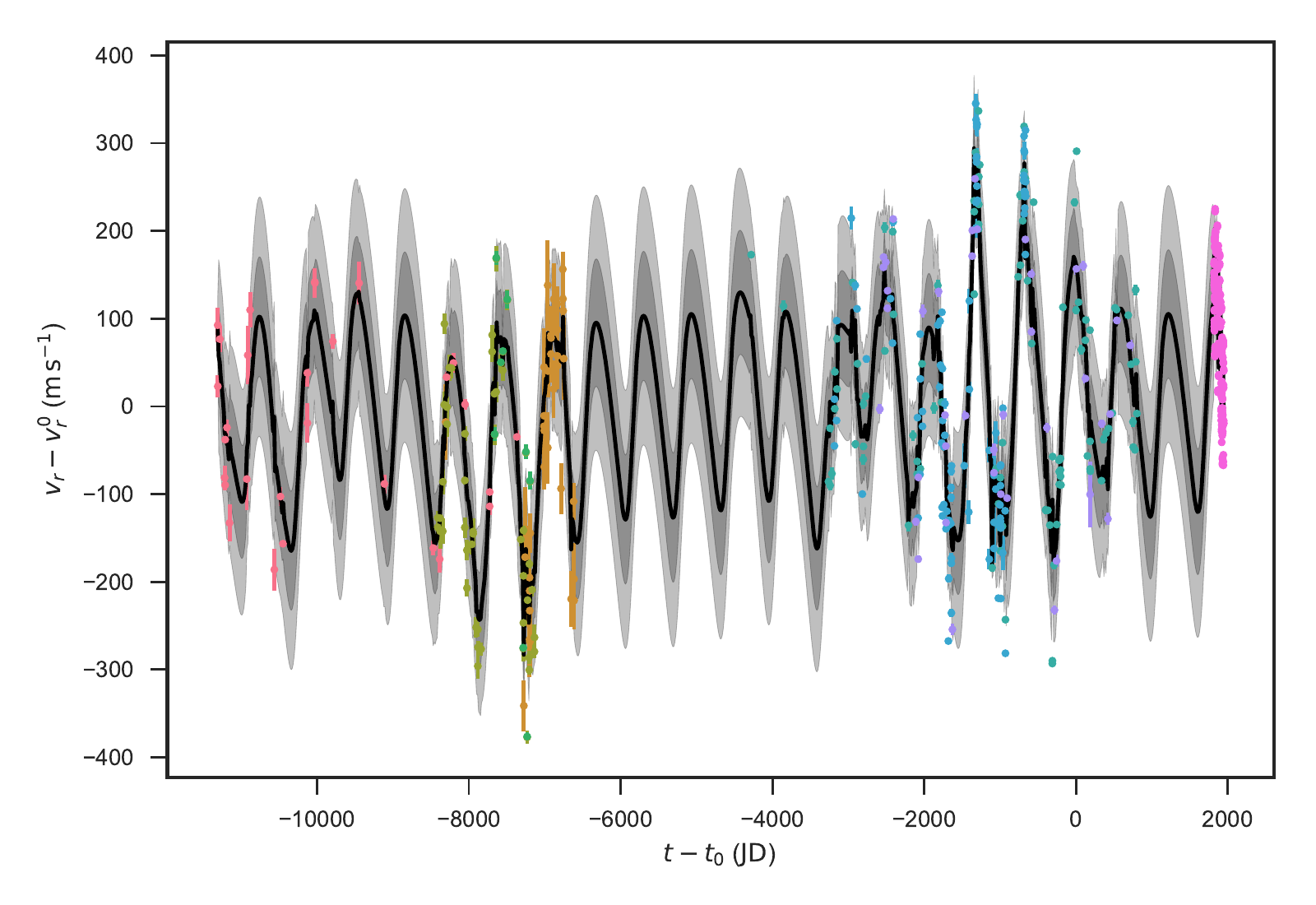}

\caption{\label{fig:data-space} A posterior draw from our model plotted with the
observed radial velocities.  Each site's radial velocity has been shifted by the
model's site-specific RV offset and errorbars have been scaled by the model's
site-specific errorbar scaling.  Data colors are as in Figure \ref{alldata}.
The black line gives the sum of the model's Keplerian component and the mean of
the inferred stochastic component; the dark and light bands give the 1- and
2-sigma uncertainty in the inferred stochastic component.  Between observations,
the uncertainty in the stochastic component increases; it reduces near
observations, as each observation provides some information about the stellar
noise that propogates to nearby times because of the temporal correlation in the
noise.  Stellar oscillations are apparent in the short-timescale wiggles in the
stochastic component between nearby observations; there is also a significant
long-period correlated component (represented in our model by an
exponentially-decaying eigenmode).}

\end{figure}

Our preferred model for the data uses one decaying mode and one oscillatory mode
in the CARMA process to capture the un-resolved superposition of many
asteroseismic modes (see Section \ref{sec:planet-star-parameters}); we have fit
models with more than one mode of each type, but find no improvement in
model-selection criteria from the expanded parameter space.  The priors we
impose in all cases on the parameters of the CARMA model, planetary RV signal,
and instrumental and site effects are given in Table \ref{tab:parameters}.  A
draw from the posterior for our canonical model in shown in data space in Figure
\ref{fig:data-space}.

The Kalman filter implementation of the CARMA likelihood is not unique.
\citet{Foreman-Mackey2017} show how a dimensional expansion can be used to
reduce the standard GP covariance matrix for this same process,  with
autocorrelation given by Eq.\ \eqref{eq:CARMA-ACF} under observations at times
$t_k$, to a banded form that can then be Cholesky-decomposed in the standard GP
likelihood function in $\mathcal{O}\left( n \right)$ time. These
`\emph{celerite}' models are formally equivalent to a Kalman CARMA model with $q
= p-1$ and differ only in their computational implementation.

\begin{deluxetable}{lll}
\tablecolumns{3}
\tablecaption{\label{tab:parameters} The parameters in our model Eq.\ \eqref{eq:RV-model} and their priors.  Site and instrument combinations are indexed by $j$.  Modes in the stochastic process are indexed by $k \in [1,p]$ and are either real (parameterised by a rate parameter $r_k$) or complex (parameterised by a frequency $f_k$ and quality factor $Q_k$).}
\tablehead{\colhead{Parameter} & \colhead{Prior} & \colhead{Description}}
\startdata
\cutinhead{Site/Instrument RV Offset and Uncertainty Scaling}
$\mu_j$ & $N\left[ \left\langle y_{j} \right\rangle, 10 \sigma_{y_{j}} \right] \left( \mu _j \right)\tablenotemark{a}$ & The site/instrument RV offset. \\
$\nu_j$ & $1/\nu_j, \quad \nu_j \in [1/2, 2]$ & The site/instrument uncertainty scale factor. \\
\cutinhead{Keplerian Parameters}
$K$ & $1/K, \quad K \in \left[ \min_j \sigma_{y_j}/100, \max_j 10 \sigma_{y_j} \right]$ & RV semi-amplitude. \\
$P$ & $1/P, \quad P \in [600, 700] \, \mathrm{d}$ & Keplerian period \\
$e$ & $2e,\quad 0 \leq e < 1$\tablenotemark{b} & Keplerian eccentricity \\
$\omega$ & $\mathrm{const}, \quad 0 \leq \omega < 2\pi$ & Longitude of pericentre \\
$\chi$ & $\mathrm{const}, \quad 0 \leq \chi < 1$ & Pericentre passage is at $t = \chi P$\\
\cutinhead{CARMA Parameters}
$A_k$ & $1/A_k, \quad A_k \in \left[\min_j \sigma_{y_j}/100, \max_j 10 \sigma_{y_j} \right]$ & RMS amplitude of stochastic mode \\
$r_k$ & $1/r_k, \quad \Re r_k \in \left[ 1/ (2T), 0.05 \, \mu\mathrm{Hz}\right], \Im r_k = 0$\tablenotemark{c} & The damping rates for real modes \\
$f_k$ & $1/f_k, \quad f_k \in \left[ 0.1, 10 \right] \mu \mathrm{Hz}$ & The frequency of oscillatory modes \\
$Q_k$ & $1/Q_k, \quad Q_k \in \left[ 1, 1000 \right]$ & The quality factor of oscillatory modes \\
\enddata
\tablenotetext{a}{$\left\langle y_j \right\rangle$ and $\sigma_{y_j}$ are the mean and standard deviation of the RV time series from site and instrument combination $j$.}
\tablenotetext{b}{We actually sample in and impose a (flat) prior on $x \equiv e \cos \omega$ and $y \equiv e \sin \omega$.}
\tablenotetext{c}{Here $T$ is the time span of the measurements over all sites.}
\end{deluxetable}

\section{Halo Photometry}
\label{halo}

The \kepler Space Telescope \citep{2010sci...327..977b}
suffered a critical reaction wheel failure in May 2013, which made it impossible to maintain a stable pointing and therefore continue its nominal mission. It was revived as \ktwo \citep{howell14}, balanced by orienting perpendicular to the Sun. This requires that \ktwo observes fields in the Ecliptic in $\sim80$~d Campaigns; Aldebaran was observed in Campaign~13.

The \kepler detector saturates for stars brighter than the $\sim11^\text{th}$ magnitude. Nevertheless, the excess flux deposited in a saturated pixel spills conservatively up and down the pixel column, such that it is possible to sum this `bleed column' for bright stars and still obtain precise photometry, such as was done for the brightest star in the nominal \kepler mission, $\theta$~Cyg \citep[$V = 4.48$;][]{guzik2011,thetacygwhite,guzik2016}. There are two main reasons why this is not possible for all bright stars in general. First, because the on-board data storage and downlink bandwidth from \kepler are limited, it is often not desirable to store and download the large number of pixels that are required for such bright stars. Second, if the bleed column for a sufficiently bright star reaches the edge of the chip, flux spills over and is not conserved, imposing a hard brightness limit that depends on the distance to the detector edge.
Collateral `smear' data, which are collected to help calibrate the photometric bias from stars sharing the same column as a target, can be used to reconstruct light curves for un-downloaded bright stars and thereby avoid bandwidth constraints \citep{k2smear}, but these data are still rendered unusable if the bleed column falls off the edge of the chip and contaminates the smear rows.

Bright stars have a wide, complicated, position-dependent point spread function (PSF) arising from diffraction and scattering from secondary and higher-order reflections inside the instrument, with the result that they may contaminate thousands of nearby pixels with significant flux. We can therefore use this `halo' of unsaturated pixels for photometry. The brightness of this halo varies in the same way as that of the primary star, and we therefore obtain data in a region of 20~pixel radius around the mean position of Aldebaran, and discard saturated pixels.
In this paper we proceed as in \citet{White2017}, in which the method was demonstrated on the seven bright Pleiades, with only minor changes.

The flux $f_i$ at each cadence $i$ is constructed as a weighted sum of pixel values $p_{ij}$:

\begin{equation}
	f_i = \sum_{j=1}^{M} w_j p_{ij}.
\end{equation}

\noindent We choose the weights $w_j$ such that they lie between 0~and~1, add to unity, and minimize the Total Variation (TV) of the weighted light curve. In the continuous case, $n$-th order TV is defined as the integral of the absolute value of the $n$-th derivative of a function; in the discrete case, replacing the derivative with finite differences, first-order TV becomes

\begin{equation}
\text{TV} = \dfrac{\sum_{i=1}^{N} |{f}_i - {f}_{i-1}|}{\sum_{j=1}^{N} {f}_j}
\end{equation}

\noindent and likewise second-order TV the equivalent expression in second-order finite differences. The efficacy of this method was recently confirmed by \citet{kallinger2017}, comparing BRITE-Constellation observations of Atlas to \ktwo halo photometry and finding excellent agreement in the frequency and amplitude of the reported oscillations.

In an improvement since \citet{White2017}, we use the \textsc{autograd} library \citep{maclaurinautograd} to calculate analytic derivatives for the TV objective function, which reduces the computational time for a single halo light curve on a commercial laptop from tens of minutes to a few seconds.

As a final step to reduce residual uncorrected systematics, we apply the \textsc{k2sc} \citep{k2sc} GP-based systematics correction code to the initial halo light curve, but the effect of this in the present case for Aldebaran is minimal.
There is somewhat higher than usual residual noise at harmonics of $46\,\mu\mathrm{Hz}$ ($4\,\mathrm{d}^{-1}$, the satellite thruster firing frequency), but this is nevertheless very small in comparison to the signal from Aldebaran, and may be ascribed to the large fraction of the pixel mask occupied by the bleed column from this extremely bright star.

\section{Stellar Modelling}\label{sec:mod}
\subsection{Stellar Models}\label{sec:stell_mod}
We used our determination of \numax and several combinations of the asteroseismic and spectroscopic parameters, along with luminosity, to estimate the fundamental stellar parameters, via fitting to stellar models. We used \textsc{MESA} models \citep{2011Paxton,2013Paxton} in conjunction with the Bayesian code \textsc{PARAM} \citep{2006dasilva, 2017Rod}.  A summary of our selected ``benchmark'' options is as follows;
\begin{itemize}
\item Heavy element partitioning from \cite{1993Grevesse}.
\item OPAL equation of state \citep{2002Rogers} along with OPAL opacities \citep{1996Iglesias}, with complementary values at low temperatures from \cite{2005Ferguson}.
\item Nuclear reaction rates from NACRE \citep{1999Angulo}.
\item The atmosphere model according to \cite{1966Kris}.
\item The mixing length theory to describe convection (we adopt a solar-calibrated parameter $\alpha_{\textrm{MLT}} =1.9657$).
\item Convective overshooting on the main sequence is set to $\alpha_{\textrm{ov}}=0.2H_{p}$, with $H_{p}$ the pressure scale height at the border of the convective core. Overshooting was applied according to the \cite{1975Maeder} step function
scheme.
\item No rotational mixing or diffusion is included.
\end{itemize}
We do not need to correct for the line-of-sight Doppler shift at the frequency precision available in our data \citep{2014MNRAS.445L..94D}.
\subsection{Additional modelling inputs}\label{sec:addmod}
In addition to the asteroseismic parameters, spectroscopically-determined temperature and metallicity values are needed. There exist multiple literature values for Aldebaran. We chose to compare a range of literature values to investigate what uncertainty these systematically-differing models produce in inferred stellar properties.

To ensure the values are self-consistent, when a literature value was chosen for temperature, we took the stellar metallicity from the same source i.e. matched pairs of temperature and metallicity. The final constraint is the stellar luminosity, which may be estimated as follows (e.g. see \citealt{pijpers2003}):
\begin{multline}
\log_{10} \frac{L}{L_{\odot}} = 4.0+
0.4 M_{{\textrm{bol}},\odot} -2.0 \log_{10} {\pi [{\textrm{mas}]}} -0.4(V-A_V + BC(V)).
\label{eqn:lum}
\end{multline}
The solar bolometric magnitude $M_{\textrm{bol},\odot}=4.73$ is taken from \cite{Torres2010}, from which we also take the polynomial expression for the bolometric correction $BC(V)$. We assume extinction $A_V$ to be zero.

The final constraint available for Aldebaran is the angular diameter of the star as measured by long baseline interferometry and lunar occultations \citep[$20.58 \pm 0.03$ mas;][]{richichi2005,1979ApJ...228L.111B,brown1979,panek1980}, combined with the \hipparcos parallax of $19.96 \pm 0.38$~pc to produce a physical radius constraint of $R_{\textrm{int}}=44.2\pm0.9\textrm{R}_{\odot}$.

\begin{table*}
	\centering
	\caption{Spectroscopic stellar parameters from each literature source, along with calculated luminosity. All temperature uncertainties assumed to be 50K, 0.2 dex in $\log{g}$, and 0.1 dex in [FeH]. For the two \protect{\cite{2012Sheffield}} results, the reason for discrepancy between the two sets of metallicity results is not discussed.}
	\label{tab:spec}
	\begin{tabular}{lllll} 
		\hline
		Spectrosocpy Source & $T_{\textrm{eff}}$ (K) & $\log{g}$ (dex) & [FeH] (dex) & Luminosity (L$_{\odot}$)\\
		\hline
		\cite{2012Sheffield}$_{a}$	&	3900	&	1.3	&	0.17	&	480\\
		\cite{2012Sheffield}$_{b}$	&	3900	&	1.3	&	0.05	&	480\\
		\cite{2011Prugniel}	&	3870 & 1.66 & -0.04 & 507\\
		\cite{2008Massarotti} & 3936 & 1 & -0.34 & 456\\
		\cite{2009Frasca} & 3850 & 0.55 & -0.1 & 526\\
		\hline
	\end{tabular}
\end{table*}

As Table \ref{tab:spec} shows, the spectroscopic parameters of Aldebaran are somewhat unclear, particularly $\log{g}$ and [FeH], which may have an impact on the recovered stellar properties when fitting to models. To explore what impact each parameter is having on the final stellar properties, multiple \textsc{PARAM} runs were performed, using different constraints. Two constraints potentially in tension were $\nu_{\textrm{max}}$ and the spectroscopically-determined $\log{g}$. $\nu_{\textrm{max}}$ has been shown to scale with the stellar $\log{g}$ \citep{Kjeldsen95, 2011A&A...530A.142B},
\begin{equation}
\frac{\nu_{\textrm{max}}}{\nu_{\textrm{max},\odot}}=\frac{\log{g}}{\log{g},\odot}\left(\frac{T_{\textrm{eff}}}{T_{\textrm{eff},\odot}}\right)^{-1/2}.
\label{eqn:numax}
\end{equation}

Using Eq \ref{eqn:numax} with the values in Table \ref{tab:spec} predicts $\nu_{\textrm{max}}$ in the range $0.5-6\mu$Hz. Reversing the equation to produce a predicted $\log{g}$ from the observed $\nu_{\textrm{max,obs}}=2.23\pm0.1\mu$Hz results in a predicted $\log{g}\sim1.2$ dex, using an assumed temperature of 3900K. The solar calibration values used here are $\log g_{\odot} / \left( \mathrm{m} \, \mathrm{s}^{-1}\right)=2.44 \, \mathrm{dex}$, $\nu_{\textrm{max},\odot}=3150\,\mu$Hz and $T_{\textrm{eff},\odot}=5777\,$K.

Table \ref{tab:res_mass} shows the results, for all modelling variations, both different inputs and different constraints. It shows that results with the addition of $\nu_{\textrm{max}}$ as a constraint exhibit in general smaller uncertainties, with or without the addition of $\log{g}$ as a constraint.

Recovering the mass without the use of asteroseismic constraints produces considerable scatter on the results ($0.96-1.5\textrm{M}_{\odot}$), whilst the use of asteroseismology brings the mass estimates into closer agreement with one another, with the exception of the very low metallicity solution of \cite{2008Massarotti}. Any systematic offset between asteroseismic masses and spectroscopic masses is sensitive to chosen reference mass, in agreement with \cite{2017North}.

We have also investigated running \textsc{PARAM} and relaxing one or more constraints: between runs with and without the luminosity constraint, the average absolute mass offset between the two sets of results was $<0.04\textrm{M}_{\odot}$ if we also ignore the $\nu_{\textrm{max}}$ constraint, and $<0.02\textrm{M}_{\odot}$ including the $\nu_{\textrm{max}}$.

\begin{table*}
	\centering

	\caption{Recovered stellar properties from \textsc{PARAM} using various constraints. Uncertainties quoted are the 68\% credible interval.}
	\label{tab:res_mass}
	\begin{tabular}{llll} 
		\hline
		Spectroscopy Source & Mass ($\textrm{M}_{\odot}$) & Radius ($\textrm{R}_{\odot}$) & Age (Gyr)\\
		\hline
		\multicolumn{4}{c}{$\nu_{\textrm{max}}$, $\log{g}$, $T_{\textrm{eff}}$, $R_\textrm{int}$, L and [FeH].} \\
		\hline
		\cite{2012Sheffield}$_{a}$ &$1.17^{+0.07}_{-0.07}$&$43.9^{+0.9}_{-0.9}$&$6.5^{+1.4}_{-1.1}$\\
		\cite{2012Sheffield}$_{b}$&$1.17^{+0.07}_{-0.07}$&$43.8^{+0.8}_{-0.9}$&$6.4^{+1.4}_{-1.1}$\\
		\cite{2011Prugniel}&$1.17^{+0.07}_{-0.07}$&$43.9^{+0.9}_{-0.9}$&$6.2^{+1.4}_{-1.1}$\\
		\cite{2008Massarotti}&$1.13^{+0.07}_{-0.07}$&$43.5^{+0.9}_{-0.9}$&$6.0^{+1.4}_{-1.1}$\\
		\cite{2009Frasca}&$1.02^{+0.04}_{-0.04}$&$44.0^{+0.8}_{-0.8}$&$10.3^{+1.5}_{-1.4}$\\
		\hline
		\multicolumn{4}{c}{$\log{g}$, $T_{\textrm{eff}}$, $R_\textrm{int}$, L and [FeH].} \\
		\hline
\cite{2012Sheffield}$_{a}$&$1.43^{+0.26}_{-0.24}$&$43.8^{+0.9}_{-0.9}$&$3.5^{+2.7}_{-1.4}$\\
\cite{2012Sheffield}$_{b}$&$1.27^{+0.24}_{-0.2}$&$43.8^{+0.9}_{-0.9}$&$4.9^{+3.6}_{-2.0}$\\
\cite{2011Prugniel}&$1.25^{+0.22}_{-0.19}$&$43.8^{+0.9}_{-0.9}$&$5.0^{+3.5}_{-2.0}$\\
\cite{2008Massarotti}&$0.95^{+0.11}_{-0.05}$&$43.8^{+0.9}_{-0.9}$&$10.2^{+2.4}_{-3.1}$\\
\cite{2009Frasca}&$0.96^{+0.04}_{-0.04}$&$44.4^{+0.8}_{-0.8}$&$11.6^{+1.4}_{-1.8}$\\
		\hline
		\multicolumn{4}{c}{$\nu_{\textrm{max}}$, $T_{\textrm{eff}}$, $R_\textrm{int}$, L and [FeH].} \\
		\hline
\cite{2012Sheffield}$_{a}$&$1.17^{+0.07}_{-0.07}$&$43.9^{+0.9}_{-0.9}$&$6.5^{+1.4}_{-1.1}$\\
\cite{2012Sheffield}$_{b}$&$1.16^{+0.07}_{-0.07}$&$43.8^{+0.8}_{-0.9}$&$6.4^{+1.4}_{-1.1}$\\
\cite{2011Prugniel}&$1.16^{+0.07}_{-0.07}$&$43.9^{+0.9}_{-0.9}$&$6.4^{+1.5}_{-1.2}$\\
\cite{2008Massarotti}&$1.13^{+0.07}_{-0.07}$&$43.5^{+0.9}_{-0.9}$&$5.9^{+1.4}_{-1.0}$\\
\cite{2009Frasca}&$1.15^{+0.07}_{-0.07}$&$43.9^{+0.9}_{-0.8}$&$6.5^{+1.5}_{-1.2}$\\
		\hline
	\end{tabular}

\end{table*}

\section{The $\gamma$~Draconis Problem is Not a Problem}
\label{gamma_dra}

$\gamma$~Draconis (also known as Eltanin, or $\gamma$~Dra) is a second-magnitude K5III giant which had been observed by \citet{hatzes1993} as part of the same campaign that led to the discovery of Aldebaran~b and Pollux~b. \citet{2018arXiv180105239H} have recently shown that the 702~d period RV variations of the putative Eltanin~b disappeared from 2011-2013, returning in 2014 with a different amplitude and phase. They ascribe this to a previously-unidentified kind of stellar variability, and warn that ``Given that the periods found in $\alpha$~Tau [Aldebaran] are comparable to those in $\gamma$~Dra and both stars are evolved with large radii, a closer scrutiny of the RV variability of $\alpha$~Tau is warranted.''

For several reasons we are not convinced that this poses an issue for Aldebaran or for K~giant RV planets more generally. \citet{2018arXiv180105239H} suggest that the new type of stellar variability may be oscillatory convective modes, but both Aldebaran and Eltanin have much lower luminosities and longer periods than would seem to be allowed by the period-luminosity relation predicted for these otherwise unobserved modes \citep[Figure~9]{2015MNRAS.452.3863S,2018arXiv180105239H}. If these are identified with the long secondary periods (LSPs) observed in some bright red giants ($L \approx 1000\, \lsun$), then Aldebaran and Eltanin are both too faint, and lack the mid-IR excess typical of LSP stars \citep{2009ApJ...707..573W}. It would also be surprising if the shape of the RV curve could reproduce the harmonic structure of an eccentric Keplerian such as in Aldebaran~b ($e = \EccentricityRange{}$), let alone the much higher eccentricities observed in other giants \citep[e.g. $\iota$~Dra~b, $e=0.7$:][]{2002ApJ...576..478F}.

It would be a cruel conspiracy of nature if red giants support a type of oscillation which is common and closely resembles a planetary signal. We believe this cannot be the case: the populations of planets around subgiants and giants evolved from intermediate mass stars are similar \citep{2014A&A...566A.113J}, as expected if subgiants evolve into giants and retain their planetary systems, and with very different stellar structures subgiants and giants are unlikely to share modes of long-period pulsation. The similarity of distributions of systems hosted by subgiant and giants could not be reproduced if a large fraction of the giants' planets were false positives. Moreover giant planets are expected to be common around intermediate-mass stars \citep{2008ApJ...673..502K}, and some are definitely known to be bona fide planets because they transit their star \citep[e.g.][]{2014A&A...562A.109L,2015A&A...573L...6O,2016AJ....152..185G,2017AJ....154..254G}.

We therefore believe that either Eltanin~b is not a false positive, or if it is, that it is not a common type of false positive, and is unlikely to affect our certainty that Aldebaran~b is real. \citet{2018arXiv180105239H} offer an alternative to the pulsation hypothesis, namely beating between the stellar rotation and the planetary signal; this does not seem implausible.

\bibliography{aldebaran}

\end{document}